\documentclass[twocolumn,showpacs,preprintnumbers,amsmath,amssymb]{revtex4}

\usepackage{graphicx}
\usepackage{dcolumn}
\usepackage{bm}
\def\overlinen#1{{\bar{#1}}}

\def\calS{{\mathcal{A}}}

\def\|{|\!|}

\newcommand{\bea}{\begin{eqnarray}}\newcommand{\eea}{\end{eqnarray}}
\newcommand{\ba}{\begin{array}}\newcommand{\ea}{\end{array}}
\newcommand{\bit}{\begin{itemize}}\newcommand{\eit}{\end{itemize}}
\newcommand{\ben}{\begin{enumerate}}\newcommand{\een}{\end{enumerate}}

\newcommand{\lf}{\left}

\newcommand{\noi}{\noindent}\newcommand{\non}{\nonumber}

\newcommand{\ran}{\rangle}
\newcommand{\ri}{\right}

\newcommand{\al}{\alpha}
\newcommand{\bt}{\beta}\newcommand{\ga}{\gamma}

\newcommand{\De}{\Delta}

\usepackage{color}
\usepackage{graphics}
\usepackage{epsf}
\newcommand{\bs}{\begin{slide}}
\newcommand{\es}{\end{slide}}
\def\boll#1{\mbox{\boldmath\footnotesize $#1$\normalsize\unboldmath}}
\def\bol#1{\mbox{\boldmath $#1$\unboldmath}}

\begin{document}

\title{'t\,Hooft's quantum determinism --- path integral viewpoint }\author{M. Blasone}
\email{blasone@sa.infn.it}
\author{P. Jizba}
\email{p.jizba@fjfi.cvut.cz}
\author{H. Kleinert}
\email{kleinert@physik.fu-berlin.de} \affiliation{
${}^{*}$Dipartimento di  Fisica, Universit\`{a} di Salerno, I-84100 Salerno, Italy\\
${}^{\dag}$FNSPE, Czech Technical University, B\v{r}ehov\'{a} 7, 115
19 Praha 1, Czech Republic\\
${}^{\ddag}$Institut f\"{u}r Theoretische Physik, Freie
Universit\"{a}t Berlin, Arnimallee 14 D-14195 Berlin, Germany}

\begin{abstract}
We present a path integral formulation of 't~Hooft's derivation of
quantum from classical physics. Our approach is based on two
concepts: Faddeev-Jackiw's treatment of constrained systems and
Gozzi's path integral formulation of classical mechanics. This treatment is compared with
 our earlier one
[quant-ph/0409021] based on Dirac-Bergmann's method.
\end{abstract}

 \maketitle

\section{Introduction}

In recent years, there has been a revival of interest in
the conceptual
foundations of quantum mechanics. In particular, a great deal of
effort has gone into the construction of deterministic theories from
which quantum mechanics would emerge. Proposals in this direction
are now of considerable topical interest as  evidenced by various
recent monographs~\cite{monographs} and this series of
workshops~\cite{elzeII}.

The usual caution toward the idea of deriving quantum from
classical physics is mainly based  on the Bell inequalities. The
fact that quantum mechanics at laboratory scales obeys these
inequalities is usually taken for granted to be true at all
scales. This perception persists even if such fundamental concepts
as rotational symmetry or isospin
--- on which the Bell inequalities are based
--- may simply cease to exist, for instance at
Planck scale. In fact, at present, no viable experiment can rule out
the possibility that quantum mechanics is only the low-energy limit
of some more fundamental underlying (possibly even non--local)
deterministic mechanism that operates at very small scales.

An interesting deterministic route to quantum physics was recently
proposed by 't~Hooft~\cite{tHooft,tHooft3,tHooft22}, motivated by
black-hole thermodynamics and the so-called {\em holographic
principle\/}~\cite{tHooft2,Bousso}. The main concept of 't~Hooft's
approach resides in {\em information loss\/}, which, when
inflicted upon a deterministic system, can reduce the physical
degrees of freedom so that quantum mechanics emerges. The
information loss together with certain accompanying non-trivial
geometric phases may explain the observed non-locality in quantum
mechanics. This idea has been further developed by several authors
\cite{BJK,tHooft22,BJV3,Elze,Halliwell:2000mv,tHooft3,BJV1}, and
it forms the basis also of this paper.

Our aim is to study 't~Hooft's quantization procedure by means of
path integrals, as done in our previous work~\cite{BJK}.  However,
in contrast to Ref.~\cite{BJK} we treat the constrained dynamics
--- the key element in 't~Hooft's method --- by means
of the Faddeev-Jackiw technique~\cite{F-J}. The constrained
dynamics enters into 't~Hooft's scheme twice: first, in the
classical starting Hamiltonian which is of first order in the
momenta and thus singular in the Dirac-Bergmann sense~\cite{Dir2}.
Second, in the information loss condition that we impose to
achieve quantization~\cite{BJK}. It is thus clear that a better
understanding of 't~Hooft's quantization scheme is closely related
to a proper treatment of the involved constrained dynamics. In our
previous paper~\cite{BJK} this has been done by means of the
customary Dirac-Bergmann technique, which is often cumbersome.
Here we want to point out the simplifications arising from the
alternative Faddeev-Jackiw method, which allows a clearer
exposition of the basic concepts.

The paper is organized as follows: In Section~II we briefly
discuss the main features of 't\,Hooft's scheme. By utilizing the
Faddeev-Jackiw procedure we present in Section~III  a Lagrangian
formulation of 't\,Hooft's system, which allows us to quantize it
via path integrals in configuration space. It is shown that the
fluctuating system produces a classical partition function. In
Section IV we make contact with Gozzi's superspace path integral
formulation of classical mechanics. In Section V we introduce
't\,Hooft's constraint which accounts for information loss. This
is again handled by means of Faddeev-Jackiw analysis. Central to
this analysis is the fact that 't~Hooft's condition breaks the
BRST symmetry and allows to recast the classical generating
functional into a form representing a genuine quantum-mechanical
partition function. A final discussion is given in Section~VI.

\section{'t Hooft's quantization procedure}

We begin with a brief review of the main aspects of 't~Hooft's
quantization procedure~\cite{tHooft22,tHooft3} to be used in this
work. The general idea is that a simple class of classical systems
can be described by means of Hilbert space techniques, although they
are fully deterministic. After imposing certain constraints
expressing information loss (or dissipation), one obtains quantum
systems. Several simple models were given by 't Hooft to illustrate
his idea, both with discrete and continuous time.

\subsection{Discrete-time version}

\begin{figure}[ht]
\begin{center}
\includegraphics*[width=3.7cm]{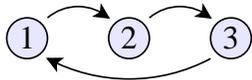}
\end{center}
\vspace{-.5cm}
\caption{Three-state universe.} \label{fig1}
\end{figure}

The simplest example \cite{tHooft3} is a three-state clock universe
with a cyclic deterministic evolution pictured in Fig.\ref{fig1}.
A Hilbert space is associated with this system consisting of the
vectors \cite{tHooft3}:
\bea
 |\psi\rangle \, =\, \al |1\rangle + \bt |2\rangle + \ga
 |3\rangle\, .
\eea
At each discrete time point $t=1,2,3,4, \dots $, the system jumps
cyclically. The time evolution may be represented by the unitary
operator
\begin{eqnarray} |\psi\ran_{t+1} \, =\, \lf(\ba{ccc}
0&0&1\\1&0&0\\0&1&0\ea\ri) |\psi\ran_t = U(t+1,t) |\psi\ran_t\, .
\end{eqnarray}
The probabilities for being in a given state are:
\begin{eqnarray} P(1)=|\al|^2 ; \quad P(2)=|\bt|^2 ; \quad
P(3)=|\ga|^2\, .
\end{eqnarray}
In a basis in which $U$ is diagonal,
it has
for a single time step  $ \Delta t=1$ the form:
\begin{eqnarray}  U(t+1,t)=\exp(-iH \De t) ; H =  \lf(\ba{ccc} 0&
&\\&-2\pi/3&\\&&2\pi/3\ea\ri)\! . \end{eqnarray}
A quantum theory can be said to be deterministic if, in the
Heisenberg picture, a complete set of operators $O_i(t)$
($i=1,..,N$) exist, such that:
\begin{eqnarray} [ O_i(t), O_j(t')] = 0, \quad \forall t, t' ; \quad i, j
=1, .. ,N\, . \end{eqnarray}
These operators are called ``be-ables'' \cite{tHooft3}.
The above three-state system is obviously deterministic in this
sense.
\begin{figure}[ht]
\begin{center}
\includegraphics*[width=4.9cm]{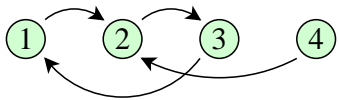}
\end{center}
\vspace{-.5cm}
\caption{Four-state universe with non-unitary
evolution.} \label{fig2}
\end{figure}
It is also possible to have systems for which the evolution is not
unitary, at least for a certain number of time steps \cite{tHooft3}.
An example is given by the system in Fig.\ref{fig2} for which the
time evolution generator is given by
\begin{eqnarray} U_d(t+1,t) \, =\, \lf(\ba{cccc}
0&0&1&0\\1&0&0&1\\0&1&0&0
\\0&0&0&0\ea\ri)\, .
\end{eqnarray}
An important concept which arises here is that of equivalence
classes \cite{tHooft3}. In our case, the states $|1)$ and $|4)$ are
equivalent, in the sense that they end up in the same state after a
finite time.

Quantum states are thus identified with equivalence classes:
\begin{eqnarray}  |1\ran\equiv \{|1),|4) \} , \quad
|2\ran\equiv \{|2)\} , \quad |3\ran\equiv \{|3)\} \, ,
\end{eqnarray}
in terms of which the time evolution becomes
unitary again.

\subsection{Continuous-time version}

Classical systems of the form
\begin{eqnarray}
 H = p_{a}\, f^{a}({\bol{q}})\, , \label{2.B.1} \end{eqnarray}
with repeated indices summed, evolve deterministically even after
quantization \cite{tHooft3}. This happens since in the Hamiltonian
equations of motion
\begin{eqnarray} \dot{q}^a &=& \{q^{a}, H\} = f^{a}({\bol{q}})\, , \non \\ [3mm]
\dot{p}_{a} &=& \{p_{a}, H\} = - p_{a}
\frac{\partial{f^{a}({\bol{q}})}}{\partial q^{a}}\, ,
\label{2.B.2}
\end{eqnarray}
the equation for the $q^a$  does not contain $p^a$, making the
$q^a$ {\em be-ables}.
%
The basic physical problem with these systems is that the
Hamiltonian is not bounded from below. This defect can be repaired
in the following way \cite{tHooft3}: Let $\rho({\bol{q}})$ be some
positive function of $q_a$ with $[\rho,H]=0$. Then we split
\begin{eqnarray}
&&H = H_{+} - H_{-} ,\nonumber \\[2mm]
&&H_{+} = \frac{1}{4\rho}\left( \rho + H\right)^{2}\, , \quad
H_{-} = \frac{1}{4 \rho}\left( \rho - H\right)^{2}\, ,
\label{2.B.3}
\end{eqnarray}
where $H_+$ and $H_-$ are positive definite operators satisfying
\bea [H_{+}, H_{-}]\, = \, [\rho , H ] =  0\, . \eea
We may now enforce a lower bound upon the Hamiltonian  by imposing
the constraint
\begin{equation}
H_{-}|\psi \ran = 0\, . \label{2.B.4}
\end{equation}
Then  the eigenvalues of $H$ in
\begin{eqnarray} H |\psi \ran = H_+ |\psi \ran = \rho |\psi
\ran                                                       ,
\end{eqnarray}
are trivially positive, and the equation of motion
\begin{eqnarray}
\frac{d}{dt} |\psi \rangle = -i H |\psi \rangle\,,
\end{eqnarray}
has only positive frequencies. If there are stable orbits with
period $T(\rho)$, then $|\psi\ran $ satisfies
\begin{eqnarray} e^{-i H T}|\psi\ran = |\psi\ran \quad; \qquad
\rho \,T(\rho) = 2 \pi n \, , \quad n \in \mathbb{Z} \, ,
\end{eqnarray}
so that the associated eigenvalues are discrete. The constraint
(\ref{2.B.4}) was motivated by 't Hooft  by information
loss~\cite{tHooft3}. We shall therefore refer to it as {\em
information loss condition\/}. Applications of the the above
quantization procedure were given in Refs.~\cite{BJV1}.

\section{Path integral quantization of 't~Hooft's system \label{SecIII}}

Consider the class of systems described by Hamiltonians of the type
(\ref{2.B.1}), and let us try to quantize them using path integrals
\cite{PI}. Because of the absence of a leading kinetic term
quadratic in the momenta $p_ a$, the system can be viewed as {\em
singular} and the ensuing quantization can be achieved through some
standard technique for quantization of constrained systems.

Particularly convenient technique is the one proposed by Faddeev and
Jackiw~\cite{F-J}. There one starts by observing that a Lagrangian
for 't~Hooft's equations of motion (\ref{2.B.2}) can be simply taken
as
\begin{eqnarray}
L({\bol{q}}, \dot{{\bol{q}}}, {\bol{p}}, \dot{\bol{p}}) ~= \
{\bol{p}}\cdot \dot{\bol{q}} - H({\bol{p}}, {\bol{q}})\, ,
\label{3.1}
\end{eqnarray}
with ${\bol{q}}$ and ${\bol{p}}$ being {\em Lagrangian variables\/}
(in contrast to  phase space variables). Note that $L$ does not
depend on $\dot{\bol{p}}$. It is easily verified that the
Euler-Lagrange equations for the Lagrangian (\ref{3.1}) indeed
coincide with the Hamiltonian equations (\ref{2.B.2}). Thus given
't~Hooft's Hamiltonian (\ref{2.B.1}) one can always construct a
first-order Lagrangian (\ref{3.1}) whose configuration space
coincides with the Hamiltonian phase space. By defining $2N$
configuration-space coordinates as
\begin{eqnarray}
&&\xi^a ~= ~p_a, \;\;\; a = 1, \ldots, N\, ,\nonumber \\
&&\xi^a ~= ~q^a, \;\;\; a = N+1, \ldots, 2N\, ,
\end{eqnarray}
the Lagrangian (\ref{3.1}) can be cast into the more expedient form,
namely
\begin{eqnarray}
L({\bol{\xi}}, \dot{\bol{\xi}}) ~= ~\mbox{$\frac{1}{2}$} \xi^a
\omega_{ab} \dot{\xi}^b - H({\bol{\xi}})\, . \label{3.26}
\end{eqnarray}
Here ${\bol{\omega}}$ is the $2N\times 2N$ symplectic matrix
\begin{eqnarray}
{\bol{\omega}}_{ab} ~= ~\lf(\ba{ll} 0&{\bol I}~\\-{\bol{I}}~&0
\ea\ri)_{ab}\, ,
\end{eqnarray}
which has an inverse ${\bol{\omega}}_{ab}^{-1} \equiv
{\bol{\omega}}^{ab}$. The equations of  motion read
\begin{eqnarray}
\dot{\xi}^a ~= ~{\bol{\omega}}^{ab} \frac{\partial
H({\bol{\xi}})}{\partial \xi^b}\, , \label{3.25}
\end{eqnarray}
indicating that there are no constraints on ${\bol{\xi}}$. Thus the
Faddeev-Jackiw procedure makes the system unconstrained, so that the
path integral quantization may proceeds in a standard way. The time
evolution amplitude is simply \cite{PI}
%
\begin{eqnarray}
&&\mbox{\hspace{-5mm}}\langle {\bol{\xi}}_2,t_2| {\bol{\xi}}_1, t_1
\rangle \! =\! {\mathcal{N}} \int_{\boll{\xi}(t_1) =
\boll{\xi}_1}^{\boll{\xi}(t_2) = \boll{\xi}_2}\!
{\mathcal{D}}{\bol{\xi}} \ \exp \left[ \frac{i}{\hbar }\!
\int_{t_1}^{t_2}\! dt~L({\bol{\xi}}, \dot{\bol{\xi}})\right],
\nonumber
\\
&&\mbox{\hspace{-5mm}} \label{frad}\end{eqnarray}
where ${\mathcal{N}}$ is some normalization factor, and the
measure can be rewritten as
\begin{equation}
{\mathcal{N}} \int_{\boll{\xi}(t_1) =
\boll{\xi}_1}^{\boll{\xi}(t_2) = \boll{\xi}_2}
{\mathcal{D}}{\bol{\xi}} = {{\mathcal{N}}} \int_{{\boll{q}}(t_1) =
{\boll{q}}_1}^{{\boll{q}}(t_2) = {\boll{q}}_2}
{\mathcal{D}}{\bol{q}} {\mathcal{D}}{{\bol{p}}} \, .
\label{@}\end{equation}
Since the Lagrangian
(\ref{3.1})
 is linear in ${\bol{p}}$,
 we may integrate these variables out and obtain
\begin{eqnarray}
&&\!\!\!\!\!\!\!\!\!\!\!\!\!\!\!\!\!\langle {\bol{q}}_2,t_2|
{\bol{q}}_1, t_1 \rangle = {{\mathcal{N}}} \int_{{\boll{q}}(t_1) =
{\boll{q}}_1}^{{\boll{q}}(t_2) = {\boll{q}}_2}
{\mathcal{D}}{\bol{q}} ~\prod_a \delta[ \dot{q}^a-f^a({\bol{q}})]\,
, \label{eg.1.2}
\end{eqnarray}
 where
$ \delta[{\bol{f}} ]\equiv \prod_t  \delta ({\bol{f}}(t))$ is the
functional version of Dirac's   $ \delta $-function. Hence the
system described by the Hamiltonian (\ref{2.B.1}) retains its
deterministic character even after quantization. The paths are
squeezed onto the classical trajectories determined by the
differential equations
 $\dot{\bol{q}} = {\bol{f}}({\bol{q}})$. The time evolution
amplitude (\ref{eg.1.2}) contains a sum  over only the classical
trajectories --- there are no quantum fluctuations driving the
system away from the classical paths, which is precisely what should
be expected from a deterministic dynamics.

The amplitude (\ref{eg.1.2}) can be brought into more intuitive form
by  utilizing the identity
\begin{eqnarray}
\delta\left[ {\bol{f}}({{\bol{q}}})  - \dot{{\bol{q}}} \right] ~=
~\delta[ {\bol{q}} - {\bol{q}}_{\rm cl}]~(\det {{M}})^{-1}\, ,
\end{eqnarray}
where ${M}$ is a functional matrix formed by the second functional
derivatives of the action ${\calS}[{\bol{\xi}}]\equiv \int
dt\,L({\bol{\xi}}, \dot{\bol{\xi}}) $\,:
\begin{eqnarray}
{{M}}_{ab}(t,t') ~= ~\left. \frac{\delta^2 {\calS}}{\delta
\xi^a(t)~\delta {\xi}^b(t')}~\right|_{{\bol{q}} = {\bol{q}}_{\rm
cl}} \, . \label{4.01}
\end{eqnarray}
The Morse index theorem  ensures that for sufficiently short
time intervals $t_2-t_1$ (before the system reaches its first
focal point), the classical solution with the initial condition
${{\bol{q}}}(t_1) = {\bol{q}}_1$ is unique.
In such a case Eq.~(\ref{eg.1.2}) can be brought to the form
\begin{eqnarray}
\langle {\bol{q}}_2,t_2| {\bol{q}}_1, t_1 \rangle &=&
{\tilde{\mathcal{N}}} \int_{{\boll{q}}(t_1) =
{\boll{q}}_1}^{{\boll{q}}(t_2) = {\boll{q}}_2}
{\mathcal{D}}{\bol{q}} ~\delta\left[{\bol{q}} - {\bol{q}}_{\rm
cl} \right]\, , \label{4.2}
\end{eqnarray}
with ${\tilde{\mathcal{N}}}\equiv {{\mathcal{N}}}/(\det {M})$.
Remarkably,
the Faddeev-Jackiw treatment bypasses completely
the discussion of
constraints, in contrast
with the conventional Dirac-Bergmann method~\cite{Dir2,Sunder}
where $2N$ (spurious) second-class primary constraints must be
introduced to deal with 't~Hooft's system, as done in \cite{BJK}.

\section{Emergent SUSY --- signature of
classicality}

We now turn to an interesting implication of the result (\ref{4.2}).
If we had started in Eq.(\ref{eg.1.2}) with an external current
\begin{eqnarray}
\tilde L({\bol{\xi}}, \dot{\bol{\xi}}) =
 L({\bol{\xi}}, \dot{\bol{\xi}}) + i\hbar {\bol{J}}\cdot {\bol{q}}\,
,
\end{eqnarray}
integrated again over ${\bol{p}}$, and took the trace over
${\bol{q}}$, we would end up with a generating functional
\begin{eqnarray}
{\mathcal{Z} }_{\rm CM}[\bol{J}] = \tilde{{\mathcal{N}}} \int
{\mathcal{D}}{\bol{q}} ~\delta[{\bol{q}}- {\bol{q}}_{\rm cl}]
 \exp\left[\int_{t_1}^{t_2} dt \ {\bol{J}}\cdot{\bol{q}}\right]\, .
\label{4.3}
\end{eqnarray}
This coincides with the path integral formulation of classical
mechanics postulated by Gozzi {\em et al}.~\cite{GozziI,GozziII}.
The same representation can be derived
from the classical limit of a
closed-time path integral for the
transition probabilities of a quantum particle in a
heat bath~\cite{PI,BJK},
The path integral
  (\ref{4.3})
has an interesting mathematical
structure. We may rewrite it as
\begin{eqnarray}
{\mathcal{Z} }_{\rm CM}[\bol{J}] ~&=& ~\tilde{{\mathcal{N}}} \int
{\mathcal{D}}{\bol{q}} ~\delta\left[ \frac{\delta {\calS}}{\delta
{\bol{q}}} \right] ~\det \left| \frac{\delta^2 {\calS} }{\delta q_a
(t) ~q_b(t')} \right|
\nonumber \\
&\times& ~\exp\left[\int_{t_1}^{t_2} dt \
{\bol{J}}\cdot{\bol{q}}\right]\, . \label{4.4}
\end{eqnarray}
By representing the delta functional in the usual way as a
functional Fourier integral
\begin{eqnarray*}
\delta\left[ \frac{\delta {\calS}}{\delta {\bol{q}}} \right] = \int
{\mathcal{D}} {\bol{\lambda}} ~\exp\left( i \int_{t_1}^{t_2} dt
~{\bol{\lambda}}(t) \frac{\delta {\calS}}{\delta {\bol{q}}(t)}
\right)\, ,
\end{eqnarray*}
and the functional determinant as a functional integral over two
real time-dependent Grassmannian {\em ghost variables\/} $c_a(t)$
and $\overlinen{c}_a(t)$,
\begin{eqnarray*}
&&\det \left| \frac{\delta^2 \calS }{\delta q^a (t) ~\delta q^b(t')}
\right|\nonumber \\
&&= \int {\mathcal{D}}{\mathbf{c}} {\mathcal{D}}
\overlinen{{\bol{c}}} ~\exp{\left[ \int_{t_1}^{t_2}dt dt' \
\bar{c}_a(t) \frac{\delta^2 {\calS} }{\delta{q^a} (t) \
\delta{q^b}(t')} ~{c_b}(t')\right]}\, ,
\end{eqnarray*}
we obtain
\begin{eqnarray}
&&\mbox{\hspace{-10mm}}{\mathcal{Z} }_{\rm CM} [\bol{J}] =  \int\!
{\mathcal{D}}{\bol{q}}{\mathcal{D}}{\bol{\lambda}}{\mathcal{D}}{\bol{c}}
{\mathcal{D}}\bar{{\bol{c}}}  \exp\left[ i {\mathcal{S}} +
\int_{t_1}^{t_2} \! dt ~{{\bol{J}}}\cdot{{\bol{q}}} \right]\! ,
\label{4.5a}
\end{eqnarray}
with the new action
\begin{eqnarray}
&&\mbox{\hspace{-7mm}}{\mathcal{S}}[{\bol{q}}, \bar{{\bol{c}}},
{\bol{c}}, {\bol{\lambda}}] \equiv ~\int_{t_1}^{t_2} dt \
{\bol{\lambda}}(t)\frac{\delta {\calS}}{\delta
{\bol{q}}(t)}\nonumber \\ &&\mbox{\hspace{-2mm}}-
i\int_{t_1}^{t_2}dt \int_{t_1}^{t_2} dt' ~\bar{c}_a(t)
\frac{\delta^2 {\calS} }{\delta{q^a} (t) ~\delta{q^b}(t')} \
{c_b}(t')\, . \label{4.5}
\end{eqnarray}
Since ${\mathcal{Z} }_{\rm CM}[\bol{J}]$ can be derived from
the classical limit of a closed-time path integral
for the transition  probability,
it comes to no
surprise that ${\mathcal{S}}$ exhibits BRST (and anti-BRST)
symmetry.
It is simple to check~\cite{BJK} that $\mathcal{S}$ does
not change under  the symmetry transformations
\begin{eqnarray}
&&\delta_{{\rm BRST\,}}  {\bol{q}} = \bar{\varepsilon} {\bol{c}}\,
,
 \;\;\; \delta_{{\rm BRST\,}}  \bar{{\bol{c}}} =
-i\bar{\varepsilon} {\bol{\lambda}}\, , \;\;\; \delta_{{\rm
BRST\,}} {\bol{c}} = 0\, , \nonumber \\&&\delta_{{\rm BRST\,}}
{\bol{\lambda}} = 0\, , \label{4.6}
\end{eqnarray}
where $\bar{\varepsilon}$ is a Grassmann-valued parameter (the
corresponding anti-BRST transformations are related to
(\ref{4.6}) by charge conjugation).
%
%
\noi
As noted in~\cite{GozziII}, the ghost fields $\bar{{\bol{c}}}$ and
${\bol{c}}$ are mandatory at the classical level as their r\^{o}le
is to cut off the fluctuations {\em perpendicular\/} to the
classical trajectories. On the formal side, $\bar{{\bol{c}}}$ and
${\bol{c}}$ may be identified with Jacobi
fields~\cite{GozziII,DeWitt}. The corresponding BRST charges
are related to Poincar\'{e}-Cartan integral
invariants~\cite{GozziIII}.

By analogy with the stochastic quantization the path integral
(\ref{4.5a}) can be rewritten in a compact form with the help of a
superfield~\cite{GozziI,Zinn-JustinII,PI}
\begin{eqnarray}
&&\mbox{\hspace{-10mm}}\Phi_a(t, \theta, \bar{\theta}) = ~q_a(t) +
i\theta c_a(t) -i\bar{\theta} \bar{c}_a(t) + i
\bar{\theta}\theta \lambda_a(t)\, , \label{3.23}
\end{eqnarray}
in which $\theta$ and $\overlinen{\theta}$ are anticommuting
coordinates extending the configuration space of ${\bol{q}}$
variables to a superspace. The latter is nothing but the degenerate
case of supersymmetric field theory in $d=1$ in the superspace
formalism of Salam and Strathdee~\cite{SS1}. In terms of superspace
variables we see that
\begin{eqnarray}
&&\!\!\!\!\!\!\!\!\!\!\!\int d\bar{\theta} d\theta ~{\calS}[{\bol{\Phi}}]\\
&&\!\!\!\!\!\!\!\!\!\!\!\!\!\!\!\!\!\mbox{\hspace{5mm}}= \int dt d\bar{\theta} d\theta \
L({\bol{q}}(t) + i\theta {\bol{c}}(t) - i \bar{\theta}
\bar{\bol{c}}(t) + i \bar{\theta}\theta \bol{\lambda}(t)
) =-i {\mathcal{S}}\nonumber
\end{eqnarray}
To obtain the last line we Taylor expanded $L$ and used the
standard integration rules for Grassmann variables. Together with
the identity ${\mathcal{D}} {\bol{\Phi}} =
{\mathcal{D}}{\bol{q}}{\mathcal{D}}{\bol{c}}{\mathcal{D}}\bar{{\bol{c}}}
{\mathcal{D}}{\bol{\lambda}}$ we may therefore express the
classical partition functions (\ref{4.3}) and (\ref{4.4}) as a
supersymmetric path integral with fully fluctuating paths in
superspace
\begin{eqnarray*}
{\mathcal{Z} }_{\rm CM}[\bol{J}]  &=&  \int\! {\mathcal{D}}
{\bol{\Phi}} \
\exp\left\{- \int d\theta d\bar{\theta} \
{\calS}[{\bol{\Phi}}](\theta, \bar{\theta}) \right\} \nonumber \\
&& \times  ~\exp\left\{\int dt d\theta d\bar{\theta} \
{\bol{\Gamma}}(t, \theta, \bar{\theta}){\bol{\Phi}}(t, \theta,
\bar{\theta})\right\}\, . \label{3.24}
\end{eqnarray*}
Here we have introduced the supercurrent ${\bol{\Gamma}}(t,
\theta, \bar{\theta}) = \bar{\theta} \theta {\bol{J}}(t)$.

Let us finally add that under rather general assumptions it is
possible to prove~\cite{BJK} that 't~Hooft's deterministic systems
are the {\em only} systems with the peculiar property that their
full quantum properties are classical in
the Gozzi {\em et al.} sense.
Among others, the latter also indicates that the
Koopman-von~Neumann operator formulation of classical
mechanics~\cite{KN1} when applied to 't~Hooft systems must agree
with their canonically quantized counterparts.

\section{Inclusion of
information loss}

As observed in Section~IIB, the Hamiltonian (\ref{2.B.1}) is not
bounded from below, and this is clearly true for any function
$f^a({\bol{q}})$. Hence, no deterministic system with dynamical
equations $\dot{q}^a = f^a({\bol{q}})$ can describe a
stable {\em quantum world\/}. To deal with this
situation we now employ 't\,Hooft's procedure of Section~II.B.
We assume that the system (\ref{2.B.1}) has $n$
conserved irreducible charges $C^i$, i.e.,
\begin{eqnarray}
\{ C^i, H \} = 0\, , \;\;\;\; i = 1, \ldots, n\, . \label{5.1}
\end{eqnarray}
Then we enforce a lower bound upon $H$,
by imposing
the condition that $H_-$
is zero on the physically accessible part of phase space.

The splitting  of $H$ into $H_-$ and $H_+$ is conserved in time
provided that $\{ H_-, H \} = \{ H_+, H \} = 0$, which is
ensured if
 $\{ H_+, H_- \} = 0$. Since the charges $C^i$ in
(\ref{5.1}) form an irreducible set,
 the Hamiltonians
 $H_+$
and $H_-$ must be functions of the charges and $H$ itself.
 There is a certain amount of
flexibility in finding $H_-$ and $H_+$.
For convenience
take the
following choice
\begin{eqnarray}
H_+ ~= ~\frac{(H + a_i C^i)^2}{4  a_i C^i} \, , \; \; H_- ~= \
\frac{(H - a_i C^i)^2}{4  a_i C^i} \, , \label{FCH}
\end{eqnarray}
where $a_i(t)$ are ${\bol{q}}$ and ${\bol{p}}$ independent. The
lower
bound is reached
 by choosing $a_i(t) C^i$ to be non-negative.
We shall select a combination of $C^i$ which is
${\bol{p}}$-independent [this condition may not necessarily be
achievable for general $f^a({\bol{q}})$].

In the Dirac-Bergmann quantization  approach used in our previous
paper \cite{BJK}, the information loss condition (\ref{2.B.4}) was a
first-class primary constraint. In the Dirac-Bergmann analysis, this
signals the presence of a gauge freedom --- the associated Lagrange
multipliers cannot be determined from dynamical equations
alone~\cite{Dir2}. The time evolution of observable quantities,
however, should not be affected by the arbitrariness of Lagrange
multipliers. To remove this superfluous freedom one must choose a
gauge. For details of this more complicated procedure see
\cite{BJK}.

In the Faddeev-Jackiw approach, Dirac's elaborate classification
of constraints to first or second class, primary or secondary is
avoided. It is therefore worthwhile to rephrase the entire
development of Ref.~\cite{BJK} once more in this approach. The
information loss condition may now be introduced by simply adding
to the Lagrangian (\ref{3.26}) a term enforcing
\begin{eqnarray}
H_-({\bol{\xi}}) ~= ~0\, , \label{5.3}
\end{eqnarray}
by means of a Lagrange multiplier:
\begin{eqnarray}
L({\bol{\xi}}, \dot{\bol{\xi}})  =  \mbox{$\frac{1}{2}$} \xi^a
{\bol{\omega}}_{ab} \dot{\xi}^b - H({\bol{\xi}}) - \eta
H_-({\bol{\xi}})\, , \label{5.2}
\end{eqnarray}
Alternatively, we shall eliminate one of $\xi^a$, say $\xi^1$, in
terms of the remaining coordinates ones, thus reducing the dynamical
variables to $2N-1$. Apart from an irrelevant total derivative, this
changes the derivative term $ \xi^a {\bol{\omega}}_{ab} \dot{\xi}^b$
to $ \xi^i {\bol{f}}_{ij}(\hat{{\bol{\xi}}})\dot{\xi}^j$, with
\begin{equation}
 {\bol{f}}_{ij}(\hat{{\bol{\xi}}})= \omega _{ij}- \left[ \omega
_{1i}\frac{\partial \xi^1}{\partial \xi^j}-(i \leftrightarrow
j)\right]. \label{@}\end{equation}
Eliminating $\xi^1$ also in the Hamiltonian $H$ we obtain a reduced
Hamiltonian $H_R(\bol{\xi})$,   so that we are left with a reduced
Lagrangian
\begin{eqnarray}
L_R(\hat{{\bol{\xi}}}, \dot{\hat{\bol{\xi}}})  ~= \
\mbox{$\frac{1}{2}$}\xi^i {\bol{f}}_{ij}(\hat{{\bol{\xi}}})
\dot{\xi}^j - H_R(\hat{{\bol{\xi}}})\, . \label{5.34}
\end{eqnarray}
At this point  one must worry about the
notorious operator-ordering problem, not knowing
in which temporal order
$\hat{{\bol{\xi}}}$ and $\dot{\hat{\bol{\xi}}}$ must be taken
in the kinetic term.
A
path integral
in which the kinetic term
is coordinate-dependent
can in general only be defined perturbatively, in which all
anharmonic terms
are treated as interactions. The partition function
is expanded in powers of expectation values
of products of these interactions which, in turn, are
expanded into integrals over all Wick contractions, the Feynman integrals.
Each contraction represents a Green function.
For a
  Lagrangian of the form (\ref{5.34}),
the contractions of two ${{{\xi}^i}}$'s contain  a Heaviside step
function, those of  one ${{{\xi}^i}}$ and one $\dot{{{\xi}^i}}$
contain a Dirac $ \delta $-function, and those of two
 $\dot{{{\xi}^i}}$'s contain a function $\dot  \delta (t-t')$.
Thus, the Feynman integrals run over products of distributions and
are mathematically undefined. Fortunately, a unique definition has
recently been found. It is enforced by the necessary physical
requirement that path integrals must be invariant under coordinate
transformations \cite{KC}.

The Lagrangian is processed further
with the help of Darboux's theorem~\cite{Darboux}. This allows us
to perform a non-canonical transformation $\xi^i \mapsto (\zeta^s,
z^r)$ which brings $L_R$ to the canonical form
\begin{eqnarray}
L_R({\bol{\zeta}, \dot{\bol{\zeta}}}, {\bol{z}}) ~= \
\mbox{$\frac{1}{2}$}\zeta^s {\bol{\omega}}_{st}\dot{\zeta}^t -
H_R'({\bol{\zeta}}, {\bol{z}})\, , \label{5.35}
\end{eqnarray}
where ${\bol{\omega}}_{st}$ is the canonical symplectic matrix in
the reduced $s$-dimensional space. Darboux's theorem ensures that
such a transformation exists at least locally. The variables $z^r$
are related to zero modes of the matrix
${\bol{f}}_{ij}(\hat{{\bol{\xi}}})$ which makes it non-invertible.
Each zero mode corresponds to a constraint of the system. In Dirac's
language these would correspond to the secondary constraints. Since
there is no $\dot z^r$ in the Lagrangian, the variable $z^r$ do not
play any dynamical r\^{o}le and can be eliminated using the
equations of motion
\begin{eqnarray}
\frac{\partial H_R'({\bol{\zeta}}, {\bol{z}})}{\partial z^r} ~= \
0\, . \label{5.36}
\end{eqnarray}
In general, $H_R'({\bol{\zeta}}, {\bol{z}})$ is a nonlinear function
of $z^{r_1}$. One now solves as many $z^{r_1}$ as possible in terms
of remaining $z$'s, which we label by $z^{r_2}$, i.e.,
\begin{eqnarray}
z^{r_1} ~= ~\varphi^{r_1}({\bol{\zeta}}, z^{r_2})\, . \label{5.4}
\end{eqnarray}
If $H_R'({\bol{\zeta}}, {\bol{z}})$ happens to be linear in
$z^{r_2}$, we obtain the constraints
\begin{eqnarray}
\varphi^{r_2}({\bol{\zeta}}) ~= ~0\, .
\end{eqnarray}
Inserting the constraints (\ref{5.4}) into (\ref{5.35}) we obtain
\begin{eqnarray}
L_R({\bol{\zeta}}, \dot{\bol{\zeta}}, {\bol{z}}) ~= \
\mbox{$\frac{1}{2}$}\zeta^s {\bol{\omega}}_{st}\dot{\zeta}^t -
H_R''({\bol{\zeta}}) - z^{r_2} \varphi^{r_2}({\bol{\zeta}})\, ,
\end{eqnarray}
with $z^{r_2}$ playing the r\^{o}le of Lagrange multipliers. We now
repeat the elimination procedure until there are no more
$z$-variables. The surviving variables represent the true physical
degrees of freedom. In the Dirac-Bergmann approach, these would span
the {\em reduced} phase space $\Gamma^*$. Use of the path integral
may now proceed along the same lines as in Ref.\cite{BJK}. In fact,
when Darboux's transformation is global it is possible to
show~\cite{BJK2} that the resultant path integral representation
coincides with the one in~\cite{BJK}.

\section{Summary}

In this paper we have presented a path-integral formulation
of 't~Hooft's quantization procedure, in the line of what done recently
in Ref.\cite{BJK}. With respect to our previous work, we have here
utilized the Faddeev-Jackiw treatment of
singular Lagrangians~\cite{F-J} which present several advantages with respect to the
usual Dirac-Bergmann method for constrained systems.

In particular,  one does not require the
Dirac-Bergmann distinction
 first and second class, primary and
secondary constraints used in \cite{BJK}. The Faddeev-Jackiw method is also
convenient in imposing 't~Hooft's information loss condition.

Although it appears that
the Faddeev-Jackiw method allows for considerable formal simplifications
of the treatment, more analysis is needed in order to compare with our previous
results of Ref.\cite{BJK}. This is object of work in progress \cite{BJK2}.

Note finally that according to analysis in Section~V, when we start
with the $N$-dimensional classical system (${\bol{q}}$ variables)
then the emergent quantum dynamics has $N-1$ dimensions
(${\bol{\zeta}}$ variables). This reduction of dimensionality
reflects the information loss. Our result supports the strong
version of the holographic principle~\cite{Bousso}, namely that the
{\em deterministic} degrees of freedom of a system scale with the
{\em bulk}, while the emergent {\em quantum} degrees of freedom
(i.e., truly observed degrees of freedom) scale with the {\em
surface}.

\acknowledgments

The authors acknowledge an instigating communication
with R.~Jackiw  and
very helpful discussions with
E.~Gozzi, J.M.~Pons, and F.~Scardigli. P.J. was financed
by
of the Ministry of Education of the Czech Republic under the
research plan MSM210000018. M.B. thanks MURST, INFN, INFM
for financial support. All of us acknowledge partial support from
the ESF
Program COSLAB.

\bibliography{apssamp}

\end{document}